\documentclass[showpacs,aps,graphicx]{revtex4}
\usepackage{graphicx}
\usepackage{amsmath}
\usepackage{amssymb}
\usepackage[titletoc]{appendix}

\begin{document}

\title{Geometric measure of quantum discord for a two-parameter class of states in a qubit-qutrit system under various
dissipative channels\footnote{Published in Quant. Inform. Proc.
\textbf{12}, 1109 - 1124 (2013). (DOI 10.1007/s11128-012-0458-8)}}

\author{Hai-Rui Wei, Bao-Cang Ren, Fu-Guo Deng\footnote{Author to whom correspondence should be addressed. Email address: fgdeng@bnu.edu.cn}}
\address{ Department of Physics, Applied Optics Beijing Area Major
Laboratory, Beijing Normal University, Beijing 100875, China}

\date{\today }

\begin{abstract}
Quantum discord, as a measure of all quantum correlations, has been
proposed as the key resource in certain quantum communication tasks
and quantum computational models without containing much
entanglement. Daki\'{c}, Vedral, and Brukner [Phys. Rev. Lett.
\textbf{105}, 190502 (2010)] introduced a geometric measure of
quantum discord (GMQD) and derived an explicit formula for any
two-qubit state. Luo and Fu [Phys. Rev. A \textbf{82}, 034302
(2010)] introduced another form of GMQD and derived an explicit
formula for arbitrary state in a bipartite quantum system. However,
the explicit analytical expression for any bipartite system was not
given. In this work, we give out the explicit analytical expressions
of the GMQD for a two-parameter class of states in a qubit-qutrit
system and study its dynamics for the states under various
dissipative channels in the first time. Our results show that all
these dynamic evolutions do not lead to a sudden vanishing of GMQD.
Quantum correlations vanish at an asymptotic time for local or
multi-local dephasing, phase-flip, and depolarizing noise channels.
However, it does not disappear even though $t\rightarrow\infty$ for
local trit-flip and local trit-phase-flip channels. Our results
maybe provide some important information for the application of GMQD
in hybrid qubit-qutrit systems in quantum information.\\

\emph{Keywords:} quantum correlations, geometric measure of quantum
discord, dynamics, a qubit-qutrit system, a two-parameter class of
states
\end{abstract}
\pacs{03.65.Yz, 03.67.Mn, 03.65.Ta }\maketitle

\section*{1. Introduction}

Quantum systems with quantum correlations, have some fundamental
applications for quantum information processing. However,
entanglement does not describe all the aspects of the quantum
correlations exhibited by a multipartite quantum system. There are
some other kinds of quantum correlations without entanglement that
are also responsible for the quantum advantages over their classical
counterparts
\cite{advantages1,advantages2,advantages3,advantages4,advantages5,advantages6,advantages7,advantages8,advantages9,advantages10}.
Aiming at capturing such correlations contained in a bipartite
quantum state, Ollivier and Zurek \cite{Discord} introduced quantum
discord as a measure of all quantum correlations in 2001. Recently,
quantum discord attracted a lot of attention. Unfortunately, quantum
discord is difficult to calculate and it is usually based on a
numerical maximization procedure. There are few analytical
expressions, including some special cases
\cite{analytical1,analytical2,analytical3,analytical4,analytical5}.
For general two-qubit mixed states, the situation is more
complicated. To avoid this difficulty, Daki\'{c} \emph{et al.}
\cite{GMQD1} introduced a geometric measure of quantum discord
(GMQD) which measures the quantum correlations of a quantum system
in a given state through the minimal Hilbert-Schmidt distance
between the given state and a zero-discord state, and they derived
an explicit formula for evaluating the GMQD for any two-qubit state
in 2010. Subsequently, Luo and Fu \cite{GMQD2} introduced another
form of GMQD and derived an explicit formula for evaluating the GMQD
for an arbitrary state in a bipartite quantum system. However, using
this way, usually, it is also difficult to calculate GMQD, and its
explicit analytical expression for any bipartite system was not
given.

Besides the characterization and quantification of quantum
correlations, another important problem is the behavior of these
correlations under the action of decoherence. Recent results of the
dynamics of the quantum correlation for a certain class of states in
a two-qubit system  under the influence of common noise channels
show that the quantum correlations vanishes at asymptotic time. It
maybe include a peculiar sudden change in behavior and is more
resistent to the action of the environment than entanglement
\cite{Dynamic1,Dynamic2,Dynamic3,Dynamic4,Dynamic5,Dynamic6}.
Quantum correlations for multi-valued quantum system is a new and
immature research area. Ali \cite{analytical5} studied the quantum
discord for a two-parameter class of states in a $2 \otimes d$
system in 2010. For a qubit-qutrit ($2\otimes3$) system, Ann
\cite{Ann} and Kan \cite{Khan} studied its entanglement dynamics
under the influence of dephasing and depolarizing channels,
respectively. Ali \emph{et al.} \cite{Ali1,Ali2} and Ramzan \emph{et
al.} \cite{Ramzan} studied its entanglement dynamics under phase
damping and amplitude damping channels. Karpat \emph{et al.}
investigated its correlation dynamics under dephasing in 2011
\cite{Karpat}. All the states concerned in previous works  are some
classes of the states in a qubit-qutrit system.

In this paper, we devote to investigate the GMQD for  a
two-parameter class of states in a qubit-qutrit system and the
dynamics of GMQD under the influence of  various dissipative
channels [i.e., both two independent (multi-local) and only one
(local) dephasing, phase-flip, bit- (trit-) flip, bit- (trit-)
phase-flip, and depolarizing channels] in the first time. Analytical
results are presented. Our results show that all these dynamic
evolutions do not lead to a sudden vanishing of GMQD. Quantum
correlations vanish at an asymptotic time for local or multi-local
dephasing, phase-flip, and depolarizing noise channels. However, it
does not disappear even though $t\rightarrow\infty$ for local
trit-flip and local trit-phase-flip channels.

\section*{2. Initial states, noise model and geometric measure of quantum discord } 


The class of states with real parameters in a hybrid qubit-qutrit
($2 \otimes 3$) quantum system \cite{Parameter} are given by
\begin{eqnarray}
\rho_{bc}(0)&=&
a\left(|02\rangle\langle02|+|12\rangle\langle12|\right)+
b(|\phi^{+}\rangle\langle\phi^{+}|+
|\phi^{-}\rangle\langle\phi^{-}|+|\psi^{+}\rangle\langle\psi^{+}|)+c|\psi^{-}\rangle\langle\psi^{-}|,
\label{eq.1}
\end{eqnarray}
where
\begin{eqnarray}
|\phi^{\pm}\rangle &=& \frac{1}{\sqrt{2}}(|00\rangle\pm|11\rangle),\nonumber\\
|\psi^{\pm}\rangle &=& \frac{1}{\sqrt{2}}(|01\rangle\pm|10\rangle),
\label{eq.2}
\end{eqnarray}
and  $a$, $b$, and $c$ are three real parameters, and they satisfy
the relation $2a+3b+c=1$. $\vert 0\rangle$ and $\vert 1\rangle$ are
the two eigenstates of a two-level quantum system (qubit) or the
eigenstates of a three-level quantum system (qutrit) with the other
eigenstate $\vert 2\rangle$. The two-parameter class of states
$\rho_{bc}(0)$ can be obtained from an arbitrary state of a $2
\otimes 3$ quantum system by means of local quantum operations and
classical communication (LOCC) \cite{Parameter}.

Let us briefly recall the quantum discord of a bipartite state
$\rho$ in a Hilbert space $H^{A}\otimes H^{B}$. The quantum discord
$\mathcal{D}(\rho^{AB})$, a measures of all quantum correlations, is
defined as the difference between the total correlation and the
classical correlation \cite{Discord,analytical4}, that is,
\begin{eqnarray}                                                   \label{eq.3}
\mathcal{D}(\rho^{AB})=I(\rho^{AB})-C(\rho^{AB}),
\end{eqnarray}
where
\begin{eqnarray}                                                   \label{eq.4}
I(\rho^{AB})=S(\rho^{A})+S(\rho^{B})-S(\rho^{AB}).
\end{eqnarray}
Here, $I(\rho^{AB})$ represents the quantum mutual information (the
total amount of correlations) of the two-subsystem state
$\rho^{AB}$. $\rho^{A(B)}=tr_{B(A)}(\rho^{AB})$ is the reduced
density matrix for the subsystem $A (B)$.
$S(\rho)=-tr(\rho\log_{2}\rho)$ is the von Neumann entropy of the
system in the state $\rho$. The other quantity, $C(\rho^{AB})$, is
interpreted \cite{Discord,Classical} as a measure of the classical
correlation of the two subsystems $AB$ in the state $\rho^{AB}$ and
it is defined as the maximal information that one can obtain, for
example, about $B$ by performing the complete measurement $\Pi_{k}$
on $H^A$,
\begin{eqnarray}                                                   \label{eq.5}
C_B(\rho^{AB})=\sup_{\Pi_{k}}\left\{S(\rho^{B})-\sum_{k}P_{k}S(\rho^{B|k})\right\},
\end{eqnarray}
where $\rho^{B|k}=\frac{1}{P_{k}}(\Pi_{k}\otimes
I_{B})\rho^{AB}(\Pi_{k}\otimes I_{B})$ is the postmeasurement state
of $B$ after obtaining the outcome $k$ on $A$ with the probability
$P_{k}=tr((\Pi_{k}\otimes I_{B})\rho^{AB}(\Pi_{k}\otimes I_{B}))$.
$\Pi_{k}$ is a set of one-dimensional projectors on $H^{A}$.

We note that  for a general mixed state,  the (one-side) classical
correlation of Eq.(\ref{eq.5}) may be asymmetric with respect to the
choice of subsystem to be measured ($C_A(\rho)\neq C_B(\rho)$), that
is, the quantum discord in Eq.(\ref{eq.3}) is not symmetric
($D_A(\rho)\neq D_B(\rho)$). However, it is known that $D_A(\rho),
D_B(\rho)\geq 0$,  and $D_A(\rho)=D_B(\rho)= 0$ if and only if
$\rho$ is a classical-quantum state, and $\rho$ has the following
expression:
\begin{eqnarray}
\rho^{AB}=\sum_{i,j}P_{ij}|i\rangle_A\langle
i|\otimes|j\rangle_B\langle j|,
\end{eqnarray}
where $P_{ij}$ is a probability distribution, and $|i\rangle_A$ and
$|j\rangle_B$ are the orthonormal bases of system $A$ and $B$,
respectively.

\textsl{GMQD}: the geometric measure of quantum discord of the state
$\rho$ is given by \cite{GMQD1}
\begin{eqnarray}                                                   \label{eq.6}
D_G(\rho)=\min_{\chi}\|\rho-\chi\|^2,
\end{eqnarray}
where the minimum is taken over the set of zero-discord states
[i.e., $D(\chi)=0$] and $\|\rho-\chi\|^2=tr(\rho-\chi)^2$ is the
square norm in the Hilbert-Schmidt space.

For any two-qubit state
\begin{eqnarray}                                                   \label{eq.7}
\rho^{AB}&=&\frac{1}{4}[I\otimes I+
\sum_{i=1}^{3}(x_{i}\sigma_{i}\otimes I+y_{i}I\otimes
\sigma_{i})+\sum_{i,j=1}^{3}(r_{ij}\sigma_{i}\otimes \sigma_{j})],
\end{eqnarray}
its GMQD is given by
\begin{eqnarray}                                                   \label{eq.8}
D_G(\rho)=\frac{1}{4}(\|X\|^2+\|R\|^2-k_{max}),
\end{eqnarray}
where $\sigma_{i}$ are the Pauli spin matrices, $X=(x_1,x_2,x_3)^T$,
$R$ is the matrix with elements $r_{ij}$, and $k_{max}$ is the
maximal eigenvalue of the matrix $K=XX^{T}+RR^{T}$.

For a general bipartite system $H^{A}\otimes H^{B}$, if we choose
orthonormal basis sets in local Hilbert-Schmidt spaces of Hermitian
operators  $\{X_i\}$ and $\{Y_j\}$ ($i=1,\cdots,d_{A}^2$ and
$j=1,\cdots,d_{B}^2$), any bipartite state can be written as
\begin{eqnarray}                                                   \label{eq.9}
\rho=\sum_{ij}c_{ij}X_i\otimes Y_j,
\end{eqnarray}
where $c_{ij}=tr(\rho X_i \otimes Y_j)$. Its GMQD can be expressed
as \cite{GMQD2}
\begin{eqnarray}                                                   \label{eq.10}
D_G(\rho)=tr(CC^T)-\max_A tr(ACC^TA^T),
\end{eqnarray}
where $C=(c_{ij})$ and the maximum is taken over all $d_A \times
d_A^2$-dimensional matrices $A=(a_{ki})$.
$a_{ki}=tr(|k\rangle\langle k|X_i)$. $\{|k\rangle\}$ is any
orthonormal base for $H^A$. Here $k=1,2,\cdots,d_A$ and
$i=1,2,\cdots,d_A^2$.


In our physical model of noise for a qubit-qutrit system (composed
of a two-level subsystem $A$ and a three-level subsystem $B$), the
two subsystems interact with their  environments independently. The
evolved states of the initial density matrix of such a system when
it is influenced by multi-local environments can be given compactly
by
\begin{eqnarray}                                        \label{eq.11}
\rho_{bc}^{AB}(t)=\sum_{i=1}^{2}\sum_{j=1}^{3}F_{j}^{B}E_{i}^{A}\rho^{AB}_{bc}(0)E_{i}^{A\dagger}F_{j}^{B
\dagger}.
\end{eqnarray}
Here, the operators $E_{i}^{A}$ and $F_{j}^{B}$ are the Kraus
operators which are used to describe the noise channels acting on
the qubit $A$ and the qutrit $B$, respectively. They satisfy the
completeness relations $\sum_i^2 E_{i}^{A\dagger}E_{i}^{A}=I$ and
$\sum_j^3 F_{j}^{B \dagger}F_{j}^{B}=I$ for all $t$.

\section*{3. Geometric measure of quantum discord for qubit-qutrit systems}

It is important to consider the behavior of quantum correlations
under the action of  decoherence. In this section, we investigate
what happens to the quantum correlations (i.e., GMQD) in a
qubit-qutrit system under common noise channels for qubit (qutrit):
dephasing, phase-flip, bit-(trit-) flip, bit-(trit-) phase-flip, and
depolarizing channels. The Kraus operators describing these channels
for a single qubit $A$ and a single qutrit $B$ are presented in
Appendix. The time-dependent parameters are defined as
$\gamma_{A}=1-e^{-t\Gamma_{A}}$ and $\gamma_{B}=1-e^{-t\Gamma_{B}}$,
$\gamma_{A}$, $\gamma_{B}\in[0,1]$ and $\Gamma_{A}$ ($\Gamma_{B}$)
denotes the decay rate of the subsystem $A$ ($B$). The two specific
environment noise situations will be considered: (i) local and (ii)
multi-local. In the case (i), only one part of a qubit-qutrit system
($S$) interacts with its environment. In the case (ii), both the two
parts of $S$ interact with their local environments, independently.

We choose 4 Hermitian matrices for each $H^A$,
\begin{eqnarray}                                                   \label{eq.12}
X_{1}&=&\frac{1}{\sqrt{2}}\left(\begin{array}{cc}
1&0\\
0&1\\
\end{array}
\right),\;\;\;\; X_{2}=\frac{1}{\sqrt{2}}\left(\begin{array}{cc}
0&1\\
1&0\\
\end{array}
\right),\;\;\;\; X_{3}=\frac{1}{\sqrt{2}}\left(\begin{array}{cc}
0&-i\\
i&0\\
\end{array}
\right),\;\;\;\; X_{4}=\frac{1}{\sqrt{2}}\left(\begin{array}{cc}
1&0\\
0&-1\\
\end{array}
\right),
\end{eqnarray}
and 9 matrices Hermitian matrices for each $H^B$,
\begin{eqnarray}                                 \label{eq.13}
Y_{1}&=&\frac{1}{\sqrt{3}}\left(\begin{array}{ccc}
1&0&0\\
0&1&0\\
0&0&1\\
\end{array}
\right),\;\;\;\; \;Y_{2}=\frac{1}{\sqrt{2}}\left(\begin{array}{ccc}
0&1&0\\
1&0&0\\
0&0&0\\
\end{array}
\right), \;\;\;\;\; Y_{3}=\frac{1}{\sqrt{2}}\left(\begin{array}{ccc}
0&-i&0\\
i&0&0\\
0&0&0\\
\end{array}
\right),\nonumber \\
 Y_{4}&=&\frac{1}{\sqrt{2}}\left(\begin{array}{ccc}
1&0&0\\
0&-1&0\\
0&0&0\\
\end{array}
\right),\;\;\;\;\;
 Y_{5}=\frac{1}{\sqrt{2}}\left(\begin{array}{ccc}
0&0&1\\
0&0&0\\
1&0&0\\
\end{array}
\right),\;\;\;\;\;
 Y_{6}=\frac{1}{\sqrt{2}}\left(\begin{array}{ccc}
0&0&-i\\
0&0&0\\
i&0&0\\
\end{array}
\right),\nonumber \\
 Y_{7}&=&\frac{1}{\sqrt{6}}\left(\begin{array}{ccc}
1&0&0\\
0&1&0\\
0&0&-2\\
\end{array}
\right),\;\;\;\;\;
 Y_{8}=\frac{1}{\sqrt{2}}\left(\begin{array}{ccc}
0&0&0\\
0&0&1\\
0&1&0\\
\end{array}
\right),\;\;\;\;\;
 Y_{9}=\frac{1}{\sqrt{2}}\left(\begin{array}{ccc}
0&0&0\\
0&0&-i\\
0&i&0\\
\end{array}
\right).
\end{eqnarray}
Any orthogonal base for $H^A$  can written as
\begin{eqnarray}                                 \label{eq.14}
|\theta_{\shortparallel}\rangle=\cos{\theta}|0\rangle+e^{i\varphi}\sin{\theta}|1\rangle,\nonumber\\
|\theta_{\bot}\rangle=\sin{\theta}|0\rangle-e^{i\varphi}\cos{\theta}|1\rangle.
\end{eqnarray}
The matrix  $A=(a_{ki})$ in Eq.(\ref{eq.10}) with the matrix
elements: $a_{ki}=tr(|k\rangle\langle k|X_i)$. Here, $\{|k\rangle\}$
is any orthonormal base for $H^A$ shown in Eq.(\ref{eq.14}), and
$X_i$ is a set of Hermitian matrices for $H^A$ shown in
Eq.(\ref{eq.12}). The matrix can be presented as
\begin{eqnarray}                                 \label{eq.15}
A=\frac{1}{\sqrt{2}}\left(\begin{array}{cccc}
1&\sin{2\theta}\cos{\varphi}&\sin{2\theta}\sin{\varphi}&\cos{2\theta}\\
1&-\sin{2\theta}\cos{\varphi}&-\sin{2\theta}\sin{\varphi}&-\cos{2\theta}\\
\end{array}
\right).
\end{eqnarray}

\textsl{(1) Channel without noise.} The coefficient elements
$c_{ij}=tr(\rho X_i \otimes Y_j)$ of the matrix $C$ in
Eq.(\ref{eq.10}) are given by
\begin{eqnarray}                                 \label{eq.16}
c_{11}&=&\frac{1}{\sqrt{6}},\nonumber\\
c_{17}&=&-\frac{2-9b-3c}{2\sqrt{3}},\nonumber\\
c_{22}&=&c_{33}=c_{44}=\frac{1}{2}(b-c),
\end{eqnarray}
and all the remaining matrix elements are zero.

By replacing the factors $A$ and $C=(c_{ij})$ in Eq.(\ref{eq.10})
with Eqs.(\ref{eq.15}) and (\ref{eq.16}), respectively (that is,
$D_G$ is the function of $\theta$ and $\varphi$), and calculating
the minimum of $D_G$ over $\theta$ and $\varphi$, we obtain the GMQD
for the qubit-qutrit systems under a channel without noise, that is
\begin{eqnarray}                                 \label{eq.17}
D_G(\rho^{AB})=\frac{1}{2}(b-c)^2.
\end{eqnarray}

\textsl{(2) Multi-local dephasing channel.} The coefficient matrix
elements $c_{ij}$ are given by
\begin{eqnarray}                                 \label{eq.18}
c_{11}&=&\frac{1}{\sqrt{6}},\nonumber\\
c_{17}&=&-\frac{1}{2\sqrt{3}}(2-9b-3c),\nonumber\\
c_{22}&=&c_{33}=\frac{1}{2}(b-c)\sqrt{(1-\gamma_{A})(1-\gamma_{B})},\nonumber\\
c_{44}&=&\frac{1}{2}(b-c),
\end{eqnarray}
and all the remaining matrix elements are zero.

\begin{figure}[!h]
\begin{center}
\includegraphics[width=10 cm,angle=0]{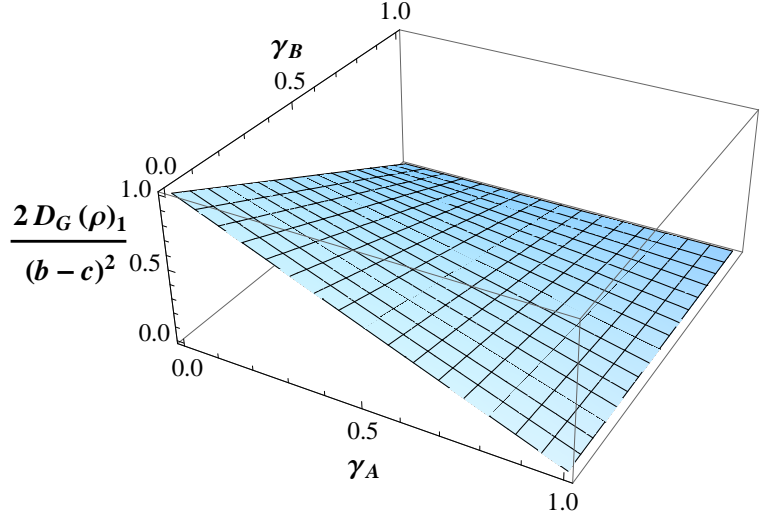}
\caption{Dynamics of GMQD for the system undergoing the multi-local
dephasing noise. $\gamma$ is the time-dependent parameter and
$\gamma_A=1-e^{-t\Gamma_A}$,
$\gamma_B=1-e^{-t\Gamma_B}$.}\label{fig1}
\end{center}
\end{figure}

By replacing the factors $A$ and $C=(c_{ij})$ in Eq.(\ref{eq.10})
with Eqs.(\ref{eq.15}) and (\ref{eq.18}), respectively, and
calculating the minimum of $D_G$ over $\theta$ and $\varphi$, we
obtain the GMQD for the qubit-qutrit systems under a multi-local
dephasing channel, that is
\begin{eqnarray}                                 \label{eq.19}
D_G(\rho^{AB})_1=\frac{1}{2}(b-c)^{2}(1-\gamma_{A})(1-\gamma_{B}).
\end{eqnarray}
Its dynamics is shown in Fig.\ref{fig1}. In the paper, we consider
the parameters $c\neq b$, that is, the initial state is  a quantum
correlation state.

\textsl{(3) Multi-local phase-flip channel.} The coefficient matrix
elements $c_{ij}$ are given by
\begin{eqnarray}                                 \label{eq.20}
c_{11}&=&\frac{1}{\sqrt{6}},\nonumber\\
c_{17}&=&-\frac{1}{2\sqrt{3}}(2-9b-3c),\nonumber\\
c_{22}&=&c_{33}=\frac{1}{2}(b-c)(1-\gamma_{A})(1-\gamma_{B}),\nonumber\\
c_{44}&=&\frac{1}{2}(b-c),
\end{eqnarray}
and all the remaining matrix elements are zero.

\begin{figure}[!h]
\begin{center}
\includegraphics[width=10 cm,angle=0]{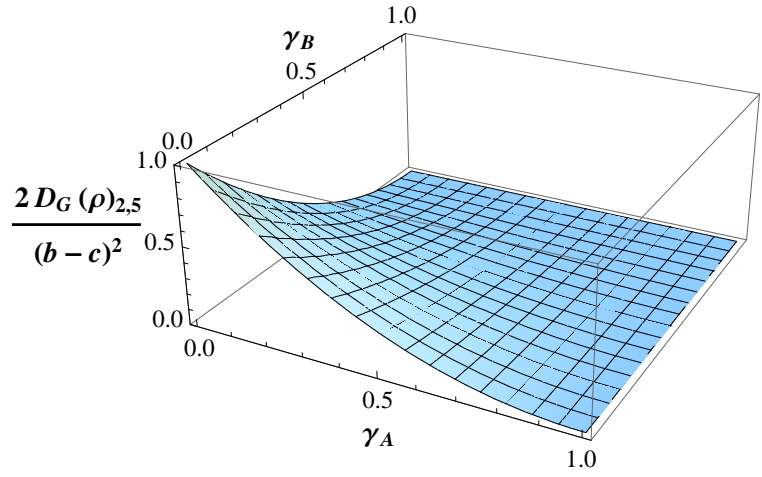}
\caption{Dynamics of GMQD for the system undergoing the multi-local
phase-flip and depolarizing noises. }\label{fig2}
\end{center}
\end{figure}

By replacing the factors $A$ and $C=(c_{ij})$ in Eq.(\ref{eq.10})
with Eqs.(\ref{eq.15}) and (\ref{eq.20}), respectively, and
calculating the minimum of $D_G$ over $\theta$ and $\varphi$, we
obtain the GMQD for the system, that is
\begin{eqnarray}                                 \label{eq.21}
D_G(\rho^{AB})_2=\frac{1}{2}(b-c)^{2}(1-\gamma_{A})^2(1-\gamma_{B})^2.
\end{eqnarray}
Its dynamics is shown in Fig.\ref{fig2}.

\textsl{(4) Multi-local bit- (trit-) flip channel.} The coefficient
matrix elements $c_{ij}$ are given by
\begin{eqnarray}                                 \label{eq.22}
c_{11}&=&\frac{1}{\sqrt{6}},\nonumber\\
c_{17}&=&\frac{1}{2\sqrt{3}}(2-9b-3c)(\gamma_{B}-1),\nonumber\\
c_{22}&=&-\frac{1}{6}(b-c)(2\gamma_{B}-3),\nonumber\\
c_{25}&=&c_{28}=\frac{1}{6}(b-c)\gamma_{B},\nonumber\\
c_{33}&=&\frac{1}{6}(b-c)(2\gamma_{B}-3)(\gamma_{A}-1),\nonumber\\
c_{36}&=&-c_{39}=\frac{1}{6}(b-c)(\gamma_{A}-1)\gamma_{B},\nonumber\\
c_{44}&=&\frac{1}{2}(b-c)(\gamma_{B}-1)(\gamma_{A}-1),
\end{eqnarray}
and all the remaining matrix elements are zero.

\begin{figure}[!h]
\begin{center}
\includegraphics[width=10 cm,angle=0]{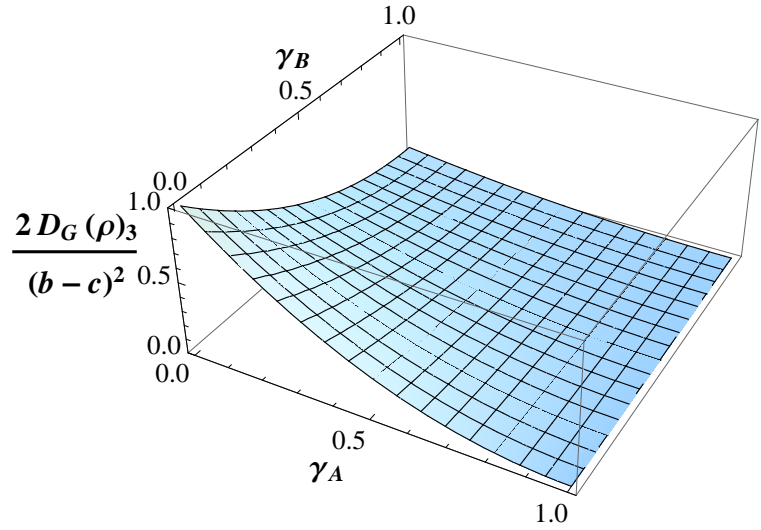}
\caption{Dynamics of GMQD for the system undergoing the multi-local
bit- (trit-) flip noise.}\label{fig3}
\end{center}
\end{figure}

By replacing the factors $A$ and $C=(c_{ij})$ in Eq.(\ref{eq.10})
with Eqs.(\ref{eq.15}) and (\ref{eq.22}), respectively, and
calculating the minimum of $D_G$ over $\theta$ and $\varphi$, we
obtain the GMQD for the system, that is
\begin{eqnarray}                                 \label{eq.23}
D_G(\rho^{AB})_3=\frac{1}{12}(b-c)^{2}(1-\gamma_{A})^2(6+5(\gamma_{B}-2)\gamma_{B}).
\end{eqnarray}
Its dynamics is shown in Fig.\ref{fig3}.

\textsl{(5) Multi-local bit- (trit-) phase-flip channel}. The
coefficient matrix elements $c_{ij}$ are given by
\begin{eqnarray}                                 \label{eq.24}
c_{11}&=&\frac{1}{\sqrt{6}},\nonumber\\
c_{17}&=&\frac{1}{2\sqrt{3}}(2-9b-3c)(\gamma_{B}-1),\nonumber\\
c_{22}&=&\frac{1}{6}(b-c)(2\gamma_{B}-3)(\gamma_{A}-1),\nonumber \\
c_{25}&=&c_{28}=\frac{1}{12}(b-c)(\gamma_{A}-1)\gamma_{B},\nonumber\\
c_{33}&=&-\frac{1}{6}(b-c)(2\gamma_{B}-3),\nonumber\\
c_{36}&=&-c_{39}=\frac{1}{12}(b-c)\gamma_{B},\nonumber\\
c_{44}&=&\frac{1}{2}(b-c)(\gamma_{B}-1)(\gamma_{A}-1),
\end{eqnarray}
and all the remaining matrix elements are zero.

\begin{figure}[!h]
\begin{center}
\includegraphics[width=10 cm,angle=0]{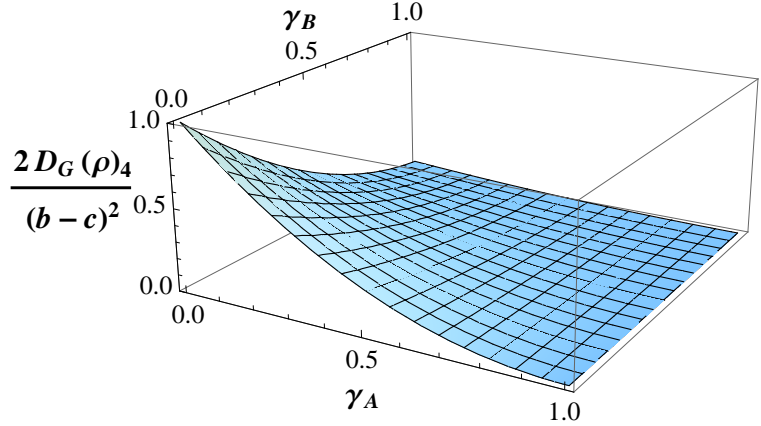}
\caption{Dynamics of GMQD for the system undergoing the multi-local
bit- (trit-) phase-flip noise.}\label{fig4}
\end{center}
\end{figure}

By replacing the factors $A$ and $C=(c_{ij})$ in Eq.(\ref{eq.10})
with Eqs.(\ref{eq.15}) and (\ref{eq.24}), respectively, and
calculating the minimum of $D_G$ over $\theta$ and $\varphi$, we
obtain the GMQD for the system, that is
\begin{eqnarray}                                                    \label{eq.25}
D_G(\rho^{AB})_4=\frac{1}{24}(b-c)^{2}(1-\gamma_{A})^2(12+\gamma_{B}(9\gamma_{B}-20)).
\end{eqnarray}
Its dynamics is shown in Fig.\ref{fig4}.

\textsl{(6) Multi-local depolarizing channel}. The coefficient
matrix elements $c_{ij}$ are given by
\begin{eqnarray}                                \label{eq.26}
c_{11}&=&\frac{1}{\sqrt{6}},\nonumber\\
c_{17}&=&\frac{1}{2\sqrt{3}}(2-9b-3c)(\gamma_{B}-1),\nonumber\\
c_{22}&=&c_{33}=c_{44}=\frac{1}{2}(b-c)(1-\gamma_{A})(1-\gamma_{B}).
\end{eqnarray}
all the remaining matrix elements are zero.

By replacing the factors $A$ and $C=(c_{ij})$ in Eq.(\ref{eq.10})
with Eqs.(\ref{eq.15}) and (\ref{eq.26}), respectively, and
calculating the minimum of $D_G$ over $\theta$ and $\varphi$, we
obtain the GMQD for the system, that is
\begin{eqnarray}                                       \label{eq.27}
D_G(\rho^{AB})_5=\frac{1}{2}(b-c)^{2}{(1-\gamma_{A})^2}{(1-\gamma_{B})^2}.
\end{eqnarray}
Its dynamics is shown in Fig.\ref{fig2}.

\begin{figure}[!h]
\begin{center}
\includegraphics[width=10 cm,angle=0]{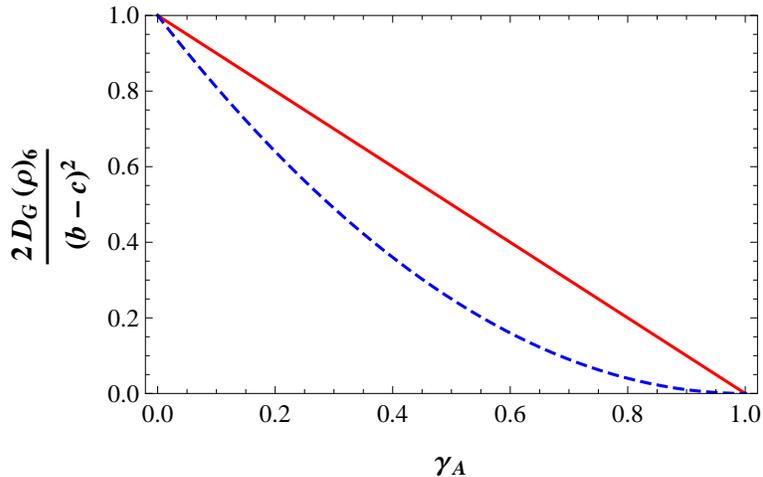}
\caption{Dynamics of GMQD for the system undergoing the various
local noises which  act on the qubit alone. The solid and
short-dashed lines correspond to dephasing and phase-flip (or
bit-flip, bit-phase-flip, depolarizing) noises,
respectively.}\label{fig5}
\end{center}
\end{figure}

\textsl{(7) Local qubit noise only}. Consider $\gamma_{B}=0$, the
GMQD can be calculated as
\begin{eqnarray}                                       \label{eq.28}
D_G^{(1)}(\rho^{AB})_6&=&\frac{1}{2}(b-c)^{2}(1-\gamma_{A}),\nonumber \\
D_G^{(2)}(\rho^{AB})_6&=&\frac{1}{2}(b-c)^{2}(1-\gamma_{A})^2.
\end{eqnarray}
Here $D_G^{(1)}(\rho^{AB})_6$ corresponds to dephasing channel, and
$D_G^{(2)}(\rho^{AB})_6$ corresponds to phase-flip, bit-flip,
bit-phase-flip or depolarizing channels. The dynamics of GMQD for
these cases are shown in Fig.\ref{fig5}

\begin{figure}[!h]
\begin{center}
\includegraphics[width=10 cm,angle=0]{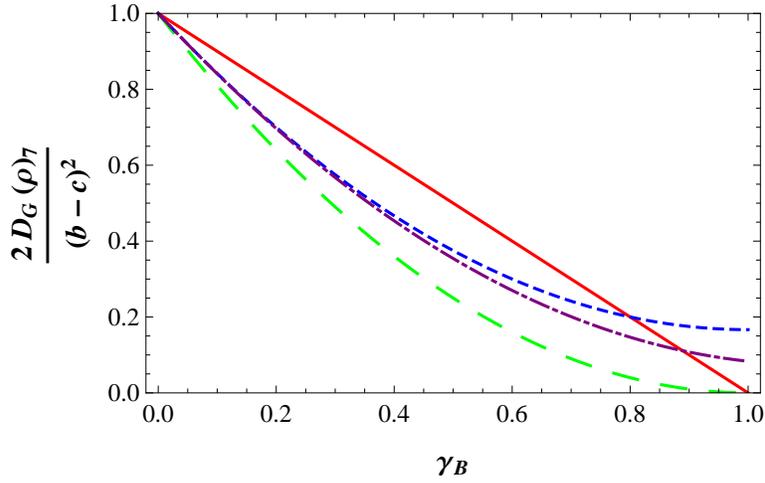}
\caption{Dynamics of GMQD for the system undergoing the various
local noises which  act on the qutrit alone. The solid, dashed,
short-dashed and dashed-dotted lines correspond to dephasing,
phase-flip ( or depolarizing), trit-flip, and trit-phase-flip
noises, respectively.}\label{fig6}
\end{center}
\end{figure}

\textsl{(8) Local qutrit noise only}. Consider $\gamma_{A}=0$, the
GMQD can be calculated as
\begin{eqnarray}                                       \label{eq.29}
D_G^{(1)}(\rho^{AB})_7&=&\frac{1}{2}(b-c)^{2}(1-\gamma_{B}),\nonumber \\
D_G^{(2)}(\rho^{AB})_7&=&\frac{1}{2}(b-c)^{2}(1-\gamma_{B})^2,\nonumber \\
D_G^{(3)}(\rho^{AB})_7&=&\frac{1}{12}(b-c)^{2}(6+5(\gamma_{B}-2)\gamma_{B}),\nonumber \\
D_G^{(4)}(\rho^{AB})_7&=&\frac{1}{24}(b-c)^{2}(12+\gamma_{B}(9\gamma_{B}-20)).
\end{eqnarray}
Here $D_G^{(1)}(\rho^{AB})_7$, $D_G^{(2)}(\rho^{AB})_7$,
$D_G^{(3)}(\rho^{AB})_7$, and $D_G^{(4)}(\rho^{AB})_7$ corresponds
to dephasing, phase-flip (or depolarizing), trit-flip, and
trit-phase-flip channels, respectively. The dynamics of GMQD for
these cases are shown in Fig.\ref{fig6}.

Eqs.(\ref{eq.28}) and (\ref{eq.29}) show that the  environment,
which  acts on the qubit alone and causes phase-flip, bit-flip,
bit-phase-flip, and depolarizing,  affects the GMQD of a hybrid
qubit-qutrit system in the same way. However, if the environment
acts on the qutrit alone, only the phase-flip and depolarizing
channels affect the GMQD of the qubit-qutrit system in the same way.

\section*{4. Discussion and Conclusion}

Quantum discord $\mathcal{D}$ for a two-parameter class of states in
a hybrid qubit-qutrit system was discussed by Ali \cite{analytical5}
in 2010. We have studied GMQD and its dynamics under the influence
of various dissipative channels, including local and multi-local
dephasing, phase-flip, bit- (trit-) flip, bit- (trit-) phase-flip,
and depolarizing channels, in the first time. Moreover, the explicit
analytical expressions were gave out. Our results show that
environment, which causes dephasing, phase-flip, bit- (trit-) flip,
bit- (trit-) phase-flip, and depolarizing of a qubit-qutrit system,
affects the quantum correlations of a hybrid qubit-qutrit system in
very different ways. All these dynamic evolutions do not lead to a
sudden vanishing of GMQD. Quantum correlations vanish at asymptotic
time  for local or multi-local dephasing, phase-flip, and
depolarizing noise channels, while it  cannot be destroyed
completely even though $t\rightarrow\infty$ for local trit-flip and
local trit-phase-flip channels. The states shown in Eq.(\ref{eq.1})
with $a=0$ (that is, $c=1-3b$) are equivalent to Werner states
\cite{Werner} in $2 \otimes 2$ system with different noise channels.
Compared with the states in two-qubit systems, their quantum
correlation all vanishes at a asymptotic time and can not occurs
quantum correlation sudden death and sudden birth under an arbitrary
Markovian dynamics \cite{Dynamic1,Dynamic3,Dynamic4}. This
phenomenon maybe provide some important information for the
application of GMQD in hybrid qubit-qutrit systems in quantum
information.

\section*{ACKNOWLEDGEMENTS}

This work is supported by the National Natural Science Foundation of
China under Grant Nos. 10974020 and 11174039,  NCET-11-0031, and the
Fundamental Research Funds for the Central Universities.

\appendix

\section{Various dissipative channels}


\textsl{Dephasing channels}: the set of Kraus operators for a single
qubit $A$ that reproduces the effect of a dephasing channel are
given  by \cite{book}
\begin{eqnarray}                                \label{eq.A1}
E_{1}^{A}=\left(\begin{array}{cc}
1&0\\
0&\sqrt{1-\gamma_{A}}\\
\end{array}
\right)\otimes I_{3}, \;\;\;\;\;\;
 E_{2}^{A}=\left(\begin{array}{cc}
0&0\\
0&\sqrt{\gamma_{A}}\\
\end{array}
\right)\otimes I_{3},
\end{eqnarray}
and those for a single qutrit $B$ can be written as \cite{Ann}
\begin{eqnarray}                          \label{eq.A2}
F_{1}^{B}=I_{2}\otimes\left(\begin{array}{ccc}
1&0&0\\
0&\sqrt{1-\gamma_{B}}&0\\
0&0&\sqrt{1-\gamma_{B}}\\
\end{array}
\right), \;\;\;\;
F_{2}^{B}=I_{2}\otimes\left(\begin{array}{ccc}
0&0&0\\
0&\sqrt{\gamma_{B}}&0\\
0&0&0\\
\end{array}
\right),\;\;\;\;
 F_{3}^{B}=I_{2}\otimes\left(\begin{array}{ccc}
0&0&0\\
0&0&0\\
0&0&\sqrt{\gamma_{B}}\\
\end{array}
\right).
\end{eqnarray}
The time-dependent parameters are defined as
$\gamma_{A}=1-e^{-t\Gamma_{A}}$ and $\gamma_{B}=1-e^{-t\Gamma_{B}}$.
Here $\gamma_{A}$, $\gamma_{B}\in[0,1]$. $\Gamma_{A}$ ($\Gamma_{B}$)
denotes the decay rate of the subsystem $A$ ($B$).


\textsl{Phase-flip channels}: the Kraus operators describing the
phase-flip channel for a single qubit $A$ are given by \cite{book}
\begin{eqnarray}                                                   \label{eq.A3}
E_{1}^{A}=\sqrt{1-\frac{\gamma_{A}}{2}}\left(\begin{array}{cc}
1&0\\
0&1\\
\end{array}
\right)\otimes I_{3},\;\;\;\;\;\;
E_{2}^{A}=\sqrt{\frac{\gamma_{A}}{2}}\left(\begin{array}{cc}
1&0\\
0&-1\\
\end{array}
\right)\otimes I_{3},
\end{eqnarray}
and those for a single qutrit $B$ can be written as
\begin{eqnarray}                                                   \label{eq.A4}
F_{1}^{B}=I_{2}\otimes\sqrt{1-\frac{2\gamma_{B}}{3}}I_3, \;
F_{2}^{B}=I_{2}\otimes\sqrt{\frac{\gamma_{B}}{3}}\left(\begin{array}{ccc}
1&0&0\\
0&e^{-i2\pi/3}&0\\
0&0&e^{i2\pi/3}\\
\end{array}
\right), \;
F_{3}^{B}=I_{2}\otimes\sqrt{\frac{\gamma_{B}}{3}}\left(\begin{array}{ccc}
1&0&0\\
0&e^{i2\pi/3}&0\\
0&0&e^{-i2\pi/3}\\
\end{array}
\right),
\end{eqnarray}
where $\gamma_{A}=1-e^{-t\Gamma_{A}}$,
$\gamma_{B}=1-e^{-t\Gamma_{B}}$, and
$\gamma_{A},\gamma_{B}\in[0,1]$. $\Gamma_{A}$ ($\Gamma_{B}$)
represents the decay rate of the subsystem $A$ ($B$).


\textsl{Bit-(Trit-) flip channels}: the Kraus operators describing
the bit-flip channel for a single qubit $A$ are given by \cite{book}
\begin{eqnarray}                                                   \label{eq.A5}
E_{1}^{A}=\sqrt{1-\frac{\gamma_{A}}{2}}\left(\begin{array}{cc}
1&0\\
0&1\\
\end{array}
\right)\otimes I_{3}, \;\;\;\;\;\;
E_{2}^{A}=\sqrt{\frac{\gamma_{A}}{2}}\left(\begin{array}{cc}
0&1\\
1&0\\
\end{array}
\right)\otimes I_{3},
\end{eqnarray}
and those for a single qutrit $B$ can be written as
\begin{eqnarray}                                                   \label{eq.A6}
F_{1}^{B}&=&I_{2}\otimes\sqrt{1-\frac{2\gamma_{B}}{3}}\left(\begin{array}{ccc}
1&0&0\\
0&1&0\\
0&0&1\\
\end{array}
\right),\;\;\;\; 
F_{2}^{B}=I_{2}\otimes\sqrt{\frac{\gamma_{B}}{3}}\left(\begin{array}{ccc}
0&0&1\\
1&0&0\\
0&1&0\\
\end{array}
\right), \;\;\;\;
F_{3}^{B}=I_{2}\otimes\sqrt{\frac{\gamma_{B}}{3}}\left(\begin{array}{ccc}
0&1&0\\
0&0&1\\
1&0&0\\
\end{array}
\right),
\end{eqnarray}
where $\gamma_{A}=1-e^{-t\Gamma_{A}}$,
$\gamma_{B}=1-e^{-t\Gamma_{B}}$, and
$\gamma_{A},\gamma_{B}\in[0,1]$.


\textsl{Bit-(Trit-) phase-flip channels}: the Kraus operators
describing the bit-phase flip channel for a single qubit $A$ are
given by \cite{book}
\begin{eqnarray}                                                   \label{eq.A7}
E_{1}^{A}=\sqrt{1-\frac{\gamma_{A}}{2}}\left(\begin{array}{cc}
1&0\\
0&1\\
\end{array}
\right)\otimes I_{3}, \;\;\;\;\;\;
E_{2}^{A}=\sqrt{\frac{\gamma_{A}}{2}}\left(\begin{array}{cc}
0&-i\\
i&0\\
\end{array}
\right)\otimes I_{3},
\end{eqnarray}
and those for a single qutrit $B$ can be written as
\begin{eqnarray}                                                   \label{eq.A8}
F_{1}^{B}&=&I_{2}\otimes\sqrt{1-\frac{2\gamma_{B}}{3}}I_3,\;\;\;\;\;\;\;\;\;\;\;\;\;\;\;\;\;\;\;
\;\;\;\;\;\;\;\;\;\; 
F_{2}^{B}=I_{2}\otimes\sqrt{\frac{\gamma_{B}}{6}}\left(\begin{array}{ccc}
0&0&e^{i2\pi/3}\\
1&0&0\\
0&e^{-i2\pi/3}&0\\
\end{array}
\right), \nonumber \\
F_{3}^{B}&=&I_{2}\otimes\sqrt{\frac{\gamma_{B}}{6}}\left(\begin{array}{ccc}
0&0&e^{-i2\pi/3}\\
1&0&0\\
0&e^{i2\pi/3}&0\\
\end{array}
\right),\;\;\;\;
F_{4}^{B}=I_{2}\otimes\sqrt{\frac{\gamma_{B}}{6}}\left(\begin{array}{ccc}
0&e^{-i2\pi/3}&0\\
0&0&e^{i2\pi/3}\\
1&0&0\\
\end{array}
\right), \nonumber \\
F_{5}^{B}&=&I_{2}\otimes\sqrt{\frac{\gamma_{B}}{6}}\left(\begin{array}{ccc}
0&e^{i2\pi/3}&0\\
0&0&e^{-i2\pi/3}\\
1&0&0\\
\end{array}
\right),
\end{eqnarray}
where $\gamma_{A}=1-e^{-t\Gamma_{A}}$,
$\gamma_{B}=1-e^{-t\Gamma_{B}}$, and
$\gamma_{A},\gamma_{B}\in[0,1]$.


\textsl{Depolarizing channels}: the set of Kraus operators that
reproduces the effect of the depolarizing channel for a single qubit
$A$  are given by \cite{book}
\begin{eqnarray}                                                   \label{eq.A9}
E_{1}^{A} = \sqrt{1-\frac{3\gamma_{A}}{4}}I_{6},\;\;\;\;
E_{2}^{A}=\sqrt{\frac{\gamma_{A}}{4}}\sigma_{1}\otimes
I_{3},\;\;\;\;
E_{3}^{A}=\sqrt{\frac{\gamma_{A}}{4}}\sigma_{2}\otimes
I_{3},\;\;\;\;
E_{4}^{A}=\sqrt{\frac{\gamma_{A}}{4}}\sigma_{3}\otimes I_{3},
\end{eqnarray}
where $\sigma_i$ ($i=1,2,3$) are the three Pauli matrices. The Kraus
operators describing a single-qutrit depolarizing noise are given by
\cite{Depolarizing}
\begin{eqnarray}                                                   \label{eq.A10}
F_{1}^{B} &=& I_{2}\otimes\sqrt{1-\frac{8\gamma_{B}}{9}}I_{3},
\;\;\;\;
F_{2}^{B} = I_{2}\otimes \frac{\sqrt{\gamma_{B}}}{3}Y,\;\;\;\;\;\;\;\;\;
F_{3}^{B} = I_{2}\otimes
\frac{\sqrt{\gamma_{B}}}{3}Z,\nonumber \\
F_{4}^{B}& =& I_{2}\otimes \frac{\sqrt{\gamma_{B}}}{3}Y^{2},\;\;\;\;\;\;\;\;\;\;\;
F_{5}^{B} = I_{2}\otimes \frac{\sqrt{\gamma_{B}}}{3}YZ,
\;\;\;\;\;\;
F_{6}^{B} = I_{2}\otimes \frac{\sqrt{\gamma_{B}}}{3}Y^{2}Z,\nonumber \\
F_{7}^{B} &=& I_{2}\otimes\frac{\sqrt{\gamma_{B}}}{3}YZ^{2},
\;\;\;\; \;\;\;\; F_{8}^{B} = I_{2}\otimes
\frac{\sqrt{\gamma_{B}}}{3}Y^{2}Z^{2},\;\;\;\;
 F_{9}^{B} =
I_{2}\otimes\frac{\sqrt{\gamma_{B}}}{3}Z^{2},
\end{eqnarray}
where
\begin{eqnarray}                                                   \label{eq.A11}
Y=\left(\begin{array}{ccc}
0&1&0\\
0&0&1\\
1&0&0\\
\end{array}
\right),\;\;\;\; Z=\left(\begin{array}{ccc}
1&0&0\\
0&e^{i2\pi/3}&0\\
0&0&e^{-i2\pi/3}\\
\end{array}
\right),
\end{eqnarray}
and $\gamma_{A}=1-e^{-t\Gamma_{A}}$,
$\gamma_{B}=1-e^{-t\Gamma_{B}}$, $\gamma_{A}, \gamma_{B}\in[0,1]$.

\end{document}